\documentclass[onecolumn]{aa}
\usepackage{graphicx}
\usepackage{txfonts}
\usepackage{deluxetable}
\usepackage{natbib}
\begin{document}
   \title{A mid-infrared study of very low mass stars and brown dwarfs in Upper Scorpius \thanks{Based on observations made with ESO Telescopes at the Paranal Observatories under program ID 075.C-0148}}

   \author{H. Bouy\inst{1,3} \thanks{Marie Curie Outgoing International Fellow}
          \and N. Hu\'elamo\inst{2,5}
          \and E.~L. Mart\'\i n\inst{3}
          \and D. Barrado y Navascu\'es \inst{4}
          \and M. Sterzik\inst{5}
          \and E. Pantin\inst{6}
          }

   \offprints{H. Bouy}

   \institute{Astronomy Department, University of California, Berkeley, CA 94720, USA\\
     \email{hbouy@astro.berkeley.edu}
     \and
     Observat\'orio Astron\'omico de Lisboa, Tapada da Ajuda, 1349-018 Lisboa, Portugal\\
     \email{nhuelamo@oal.ul.pt}
     \and 
     Instituto de Astrof\'\i sica de Canarias, C/ V\'\i a L\'actea, s/n, E-38200 - La Laguna, Tenerife, Spain\\
     \email{ege@iac.es}
     \and
     Laboratorio de Astrof\'\i sca Espacial y F\'\i sica Fundamental (LAEFF-INTA), Apdo 50727, 28080 Madrid, Spain\\
     \email{barrado@laeff.inta.es}
     \and
     European Southern Observatory,  Casilla 19001, Santiago 19, Chile\\
     \email{msterzik@eso.org}
     \and
     Services d'Astrophysique Saclay CEA F-91191 Gif-sur-Yvette France\\
     \email{pantin@sapvxg.saclay.cea.fr}
   }

   \date{Received ...; accepted 08/2006}

  \abstract
   {To investigate the formation of sub-stellar objects, we observed a sample of ultracool dwarfs members of the Upper Scorpius OB association.}
   {The properties of disks, such as their composition, life-time, frequency, and how it compares to the accretor frequency, provide important clues on the mechanisms responsible for the formation of stellar and sub-stellar objects.}
   {We report the results of mid-IR observations with VISIR at the VLT of 10 ultracool dwarfs members of the nearby Upper Scorpius OB association in four filters ranging between 8.59 (PAH1) to 12.8~$\mu$m (Ne~II), and one brown dwarf with Spitzer between 3.6 and 24~$\mu$m.}
   {Seven of our targets are detected in at least one of the bands, and we derive upper limits on the fluxes of the remaining 4. These results combined with previous studies from the literature lead to an improved disk frequency of 50$\pm$12\%. This frequency is significantly higher than that of accretors (16.3\%$\pm$6.2\%). Only one object showing mid-IR excess also has H$\alpha$ emission at a level indicating that it must be accreting. Four of the detected targets are multiple system candidates.}
   {The observed disk frequency for sub-stellar objects in the Upper Scorpius association is similar to that of stars, consistent with a common formation scenario. It is also similar to the disk fractions observed in younger clusters, suggesting that the disk lifetimes might be longer for ultracool dwarfs than for higher-mass stars. }


   \maketitle
%

\section{Introduction}
Over the last ten years, a large number of brown dwarfs (BD) and very low mass stars have been discovered in the field, in star forming regions and in clusters. One of the most debated questions is whether sub-stellar objects form the same way as stars, via the fragmentation and collapse of a molecular cloud, or as aborted stellar embryos ejected from their parent proto-stellar cluster. A comparison of stellar and sub-stellar disk properties directly addresses this question and provides crucial informations and straightforward constraints on the models of formation. The first evidences showing that brown dwarfs have disks have been given several years ago \citep{1999ApJ...525..440L,1998A&A...335..522C} and since then substantial fractions of young brown dwarfs in star forming regions have been reported to have circumstellar material \citep[see e.g.][]{2005ApJ...635L..93L,2005ApJ...620L..51L,2005ApJ...631L..69L,2005ApJ...626..498M, 2004A&A...427..245S,2003AJ....126.1515J,2003ApJ...585..372L,2001ApJ...558L..51M, 2001A&A...376L..22N} and to undergo an accretion phase \citep[e.g ][]{2002ApJ...578L.141J, 2004A&A...424..603N,2004ApJ...604..284B,2005ApJ...626..498M,2005ApJ...625..906M}. From their near-IR observations,  \citet{2003AJ....126.1515J} estimated an initial frequency of disk around BD and very low mass stars of 50\%$\pm$25\% in various star forming regions. Using Spitzer to study the population of the IC348 young cluster, \citet{2006AJ....131.1574L} estimated disk fractions of 47\% for K6--M2 stars, and 28\% for M2--M6 stars, while \citet{2005ApJ...631L..69L} reported 42\% for members later than M6. Although these results seem to show that BD originate like stars more than like ejected or evaporated embryos, our understanding of the formation processes is far from complete. It is still unclear whether one mechanism eventually prevails over the other, and how the contribution of each mechanism to the overall final population depends on the environment and initial conditions. 

The Upper Scorpius OB association (hereafter USco) is an ideal place in which to study the properties of ultracool dwarf disks. It is close and young enough \citep[145~pc and 5~Myr,][]{1999AJ....117..354D} that even its sub-stellar members have relatively bright luminosities, and the sample of known sub-stellar members is one of the largest for that type of association. Moreover, by comparing the disk properties of the 5~Myr USco to those of younger associations such as IC348 \citep{2006AJ....131.1574L,2005ApJ...631L..69L} or Cha I \citep{2005ApJ...631L..69L}, it is possible to study the disk evolution and lifetimes. We have obtained mid-infrared (mid-IR) photometry of 10 ultracool dwarf members of the association using VISIR on the VLT, and retrieved mid-IR photometry of an additional USco brown dwarf from the Spitzer public archive. In section \ref{observations}, we describe the observations, data reduction and analysis, and in section \ref{analysis}, we discuss the derived disk frequency and compare to previous studies.


\section{Observations and data reduction \label{observations}}

\subsection{Sample selection}
The original sample consists of 11 ultracool dwarfs with H$\alpha$ emission in USco, reported by \citet{2004AJ....127..449M}. Table \ref{targets} gives a summary of the targets properties. All have been spectroscopically confirmed as ultracool dwarfs and as member of the Upper Sco association. Two of the targets were classified as accretors by \citet{2004AJ....127..449M} on the basis of their H$\alpha$ emission and the saturation criterion defined by \citet{2003AJ....126.2997B} (see Table \ref{bindisk}). Four of the targets have recently been resolved as multiple system candidates by \citet{2006A&A...451..177B} and \citet{2005ApJ...633L..41L}. One of them (DENIS161833) has been confirmed as a physical pair by the two latter authors, while second epoch observations are on-going to confirm the multiplicity of the other ones.

\begin{deluxetable}{lllc}
\tablecaption{List of Targets \label{targets}}
\tablewidth{0pt}
\tablehead{
\colhead{Target} & \colhead{RA}      & \colhead{Dec}      & \colhead{SpT} \\
\colhead{}       & \colhead{(J2000)} & \colhead{(J2000)}  &                
}
\startdata
DENIS155556      & 15 55 56.0        & -20 45 18.5        & M6.5               \\
DENIS160334      & 16 03 34.7        & -18 29 30.4        & M5.5                \\
DENIS160514      & 16 05 14.0        & -24 06 52.6        & M6.0     \\
DENIS160603      & 16 06 03.9        & -20 56 44.6        & M7.5   \\
DENIS160951      & 16 09 51.1        & -27 22 42.2        & M6.0    \\
DENIS160958AB\tablenotemark{1}      & 16 09 58.5        & -23 45 18.6  & M6.5     \\
DENIS161050      & 16 10 50.0        & -22 12 51.8        & M5.5  \\
DENIS161816AB\tablenotemark{1}     & 16 18 16.2        & -26 19 08.1        & M5.5    \\
DENIS161833AB\tablenotemark{1}      & 16 18 33.2        & -25 17 50.4        & M6.0  \\
DENIS161939AB\tablenotemark{1}      & 16 19 39.8        & -21 45 35.1        & M7.0 \\
DENIS162041\tablenotemark{2}      & 16 20 41.5        & -24 25 49.0        & M7.5 \\
\enddata
\tablenotetext{1}{ Multiple systems \citep{2006A&A...451..177B,2005ApJ...633L..41L} }
\tablenotetext{2}{ Observed with Spitzer rather than VISIR}
\tablecomments{Coordinates and spectral types \citet{2004AJ....127..449M}}
\end{deluxetable}

\begin{deluxetable}{lccl}
\tablecaption{Disk, accretion and multiplicity of the Targets\label{bindisk}}
\tablewidth{0pt}
\tablehead{
\colhead{Target} & \colhead{Mid-IR} & \colhead{Accretion} & \colhead{Companion} \\
\colhead{}       & \colhead{Excess} & \colhead{}          & \colhead{}
}
\startdata
DENIS-PJ155556   & yes      & no  & no \\
DENIS-PJ160334   & yes      & no  & yes? \tablenotemark{ a} \\
DENIS-PJ160514   & \nodata  & no  & no \\
DENIS-PJ160603   & \nodata  & yes & no \\
DENIS-PJ160951   & \nodata  & no  & no \\
DENIS-PJ160958   & no       & no  & yes (12~AU) \\
DENIS-PJ161050   & no       & no  & yes? \tablenotemark{ a} \\
DENIS-PJ161816   & no       & no  & yes (21~AU) \\
DENIS-PJ161833   & no       & no  & yes (134~AU) \\
DENIS-PJ161939   & yes      & yes & yes (26~AU) \\
DENIS-PJ162041   & no\tablenotemark{ b}       & no  & no \\
\enddata
\tablenotetext{a}{As discussed in the text, we infer the possible multiplicity of DENIS-PJ160334.7-182930 and  DENIS-PJ161050.0-221252 from their SEDs, but these were not resolved with adaptive optics.}
\tablenotetext{b}{Observed with Spitzer rather than VISIR}
\tablecomments{Accretion diagnostic from \citet{2004AJ....127..449M}. No companion means no companion resolved at separations $\ge$5~AU and $\Delta$Mag(Ks)$<$2~mag \citep[][ in prep.]{2006A&A...451..177B,bouy_in_prep}.}
\end{deluxetable}

\subsection{VISIR observations}
Our mid-IR observations were obtained with VISIR, the mid-IR instrument mounted on the Cassegrain focus of Melipal (UT3)  at the VLT \citep{2004Msngr.117...12L}. The selected pixel-scale, 0\farcs075, results in a total field of view of 19\farcm2$\times$19\farcm2 on its $256\times256$ BIB detector. The observations were performed using standard chopping and nodding techniques in order to suppress the sky emission.  The chopping throws were selected between 8--10\arcsec. A standard star was observed before and/or after each target with the same configuration and at a similar airmass. The targets have been observed with the PAH1, PAH2, SIV and NeII filters, where the instrument has its best sensitivity. The effective wavelengths of these filters are described in Table \ref{filters}.

\begin{figure}
   \centering
   \includegraphics[width=0.5\textwidth]{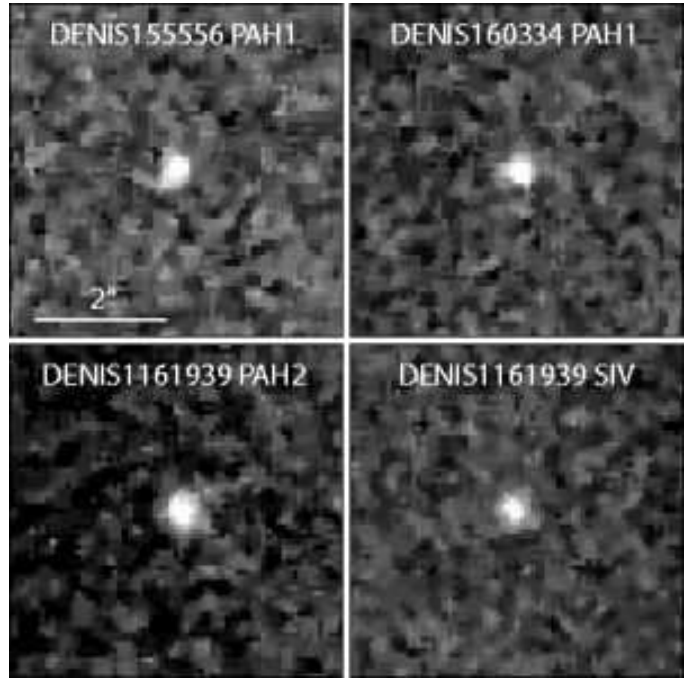}
      \caption{VISIR image stamps of DENIS155556, DENIS160334 and DENIS161939. The scale and the filter are indicated.
              }
         \label{images}
   \end{figure}

The data were processed with a customary pipeline based on IDL (Interactive Data Language) routines. The pipeline includes bad pixel correction and the shift and add of the individual frames, each of them corresponding to a chopping position \citep{2005Msngr.119...25}. Table~\ref{obs} shows a summary of the observations. Figure \ref{images} shows a sample of four of the images we obtained.

\begin{deluxetable}{lcc}
\tablecaption{Effective wavelengths and bandwidths of the VISIR Filters used for this study\label{filters}}
\tablewidth{0pt}
\tablehead{
\colhead{Filter} & \colhead{$\lambda_{0}$ [$\mu$m]} & \colhead{$\Delta$$\lambda$ [$\mu$m]}
}
\startdata
PAH1  & 8.59  & 0.42 \\
SIV   & 10.49 & 0.16 \\
PAH2  & 11.26 & 0.59 \\
NeII  & 12.81 & 0.21 \\
\enddata
\end{deluxetable}

\begin{deluxetable}{lccc}
\tablecaption{Observation Log \label{obs}}
\tablewidth{0pt}
\tablehead{
\colhead{Target} & \colhead{Filter} & \colhead{Exp. Time} & \colhead{Calibrator} \\
\colhead{}       & \colhead{}       & \colhead{[s]}       & \colhead{} 
}
\startdata
\multicolumn{4}{c}{30/05/2005}\\
\hline
DENIS161816 & PAH1 & 3600 & HD145897 \\
DENIS161833 & PAH2 & 3600 & HD145897 \\
DENIS160334 & PAH2 & 3600 & HD168592 \\
DENIS160514 & PAH2 & 3600 & HD168592 \\
DENIS160958 & PAH2 & 3600 & HD152880 \\
\hline
\multicolumn{4}{c}{31/05/2005}\\
\hline
DENIS161816 & PAH2 & 3600 & HD145897 \\
DENIS161939 & PAH2 & 3600 & HD145897 \\
DENIS160603 & PAH2 & 3600 & HD146791 \\
DENIS160951 & PAH2 & 3600 & HD146791 \\
DENIS161050 & PAH2 & 3600 & HD152880 \\
DENIS155556 & PAH2 & 3600 & HD152880 \\
\hline
\multicolumn{4}{c}{01/06/2005}\\
\hline
DENIS161939 & PAH1 & 3600 & HD151249 \\
DENIS161939 & SIV & 3600  & HD151249 \\
DENIS160334 & SIV & 3600  & HD179886 \\
DENIS160334 & PAH1 & 3600 & HD156277 \\
DENIS160603 & PAH2 & 3600 & HD138538 \\
\hline
\multicolumn{4}{c}{02/06/2005}\\
\hline
DENIS155556 & PAH1 & 3600 & HD133774 \\
DENIS155556 & SIV & 3600  & HD133774 \\
DENIS161939 & NEII & 1800 & HD152880 \\
\enddata
\end{deluxetable}

   \begin{figure}
   \centering
   \includegraphics[width=0.5\textwidth]{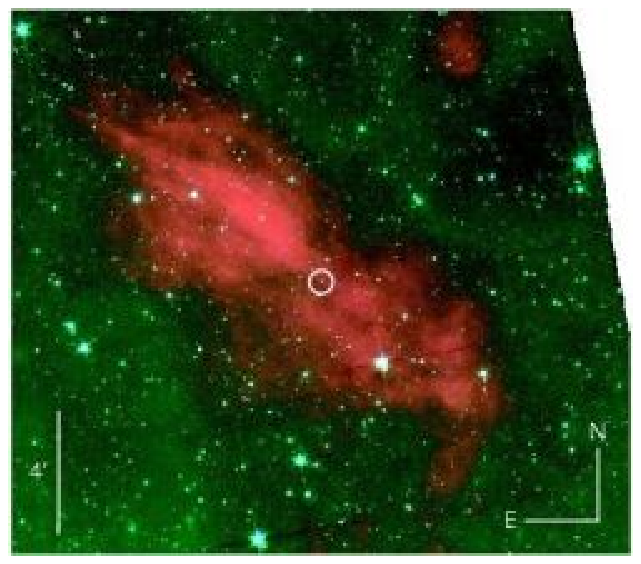}
      \caption{Three color image obtained using the Spitzer IRAC images at 3.6, 4.5 and 8.0~$\mu$m (blue, green, red respectively). The scale and orientation are indicated. DENIS162041 is indicated with a white circle. It is located in a region of higher extinction.
              }
         \label{denis1620_spitzer}
   \end{figure}

\subsection{Spitzer archival data}
In order to complement our ground-based survey, we searched the Spitzer public archive to look for observations of USco brown dwarfs. Only one object of \citet{2004AJ....127..449M} sample, the M7.5 brown dwarf DENIS-PJ162041.5-242549.0, had been observed in the course of a public survey. It fell in both the IRAC and MIPS pointings of the ``C2d continued'' legacy survey (Program 177, P.I. N. Evans). The object had been observed twice in each of the IRAC and MIPS bands, and is detected in the IRAC images only (see Figure \ref{denis1620_spitzer}).

\subsection{Analysis of the data}
\begin{figure}
   \centering
   \includegraphics[width=0.9\textwidth]{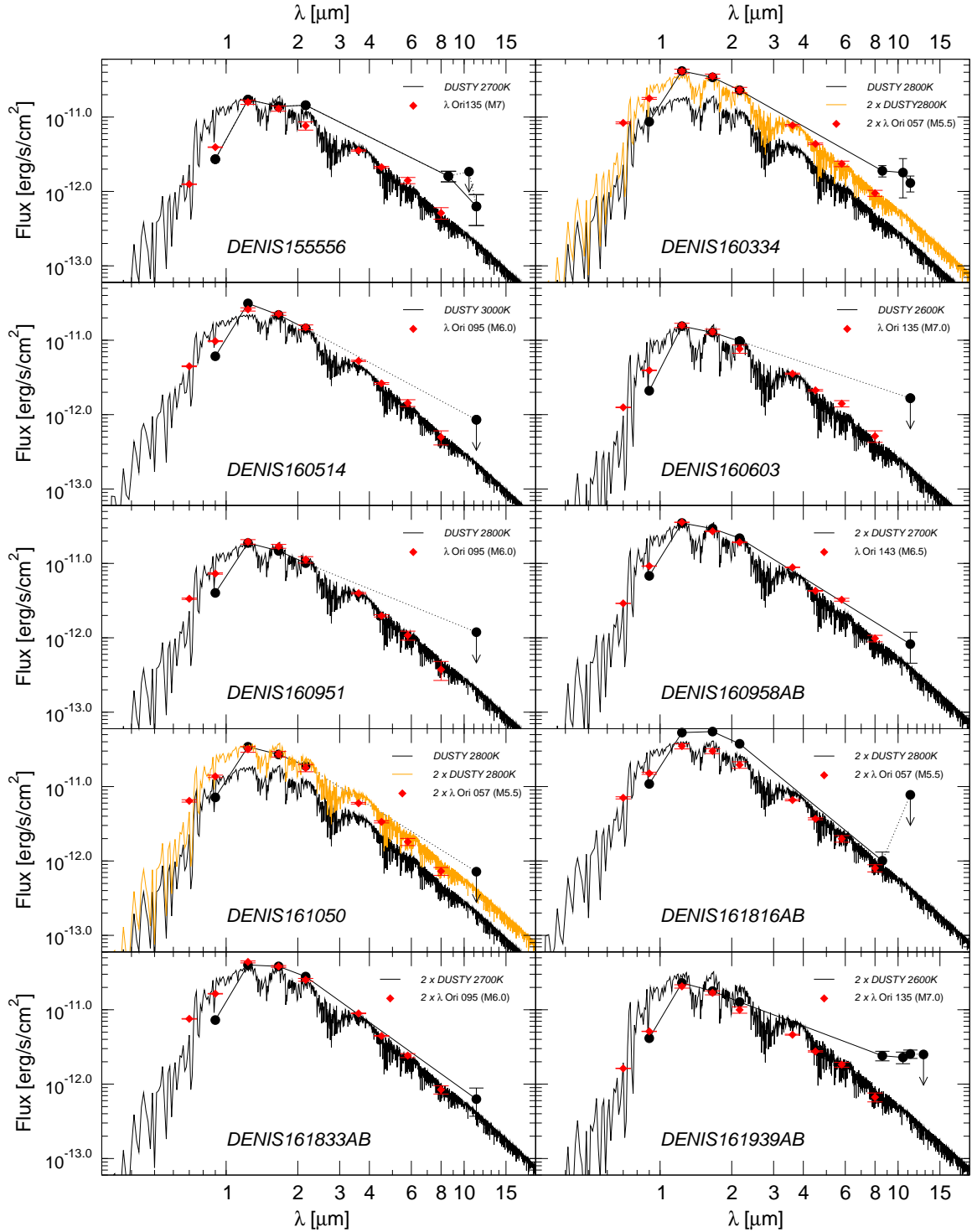}
      \caption{Spectral Energy Distributions of the 10 VISIR targets (black circles). Upper limits are indicated with arrows and connected with dotted lines. DUSTY atmospheric models \citep[with the indicated T$_{eff}$ and $log(g)$=4.0, ][]{2001ApJ...556..357A} and SEDs of $\lambda$-Ori M dwarfs (red diamonds) of similar spectral types from \citet{barrado_inprep} corrected for the distance of the association are over-plotted for comparison. The T$_{eff}$ were chosen according to \citet{2004ApJ...609..885M} Spectral Type vs T$_{eff}$ scale for USco objects. The individual components of the 4 known binary systems having similar masses \citep[see][]{2006A&A...451..177B}, their SEDs are compared to the sum of two atmospheric models/$\lambda$-Ori M dwarfs of the same effective temperature. Two objects (DENIS161050 and DENIS160334) are twice brighter than the atmospheric models corresponding to their effective temperatures. The orange curves show the comparison with the sum of two atmospheric models of the same temperature, suggesting these targets might be unresolved binaries (see discussion in the text).
              }
         \label{seds}
   \end{figure}

Most of the VISIR detections are so faint that only the central overlay of the two positive beams can be used for photometry. Aperture photometry was performed using standard procedures under IDL with the APPHOT library. An optimized aperture was derived using the curve of growth method. Since the targets are faint, the best aperture was chosen to maximize the signal-to-noise ratio on the standard stars curves of growth. The derived apertures were then used to measure the count-rate of each associated target. Uncertainties were estimated measuring the background noise weighted over the aperture. The count rates were then translated into fluxes using the calibrated fluxes of the standard stars \citep[][]{1992AJ....104.1650C, 1995AJ....110..275C,1996AJ....112..241C, 1995MNRAS.276..715C} provided by the VISIR team. The results are shown in Table \ref{photom}, and in Figure \ref{seds}.

\begin{figure}
   \centering
   \includegraphics[width=0.7\textwidth]{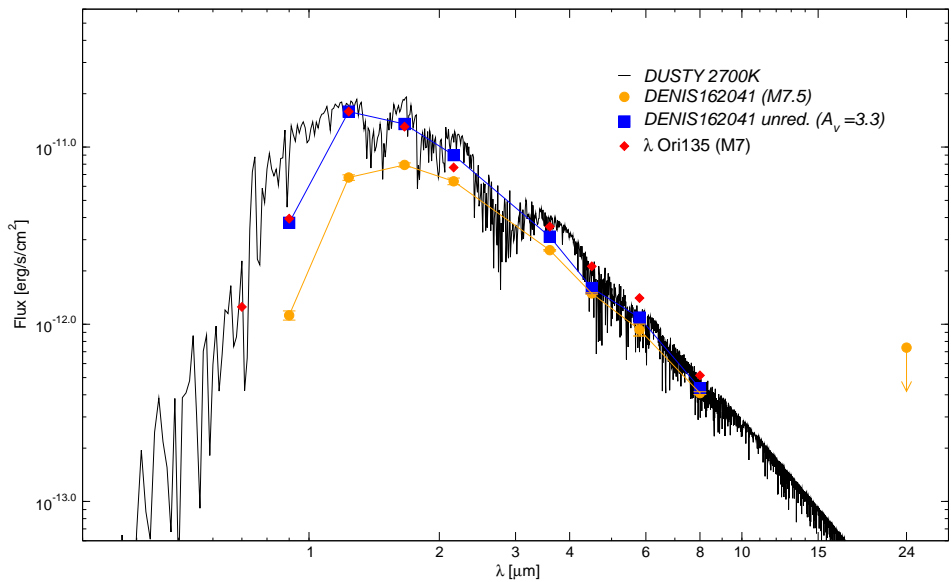}
      \caption{Spectral Energy Distributions of DENIS162041. The SED of DENIS162041 (orange) does not fit the corresponding model (black line) or comparison M7 dwarf (red diamonds). The deviation is increasing towards bluer wavelengths, suggesting that the object suffers from extinction. The SED fits relatively well the model/comparison M7 dwarf after unreddening with $A_{V}$=3.3 (blue squares). The object does not shows any excess in the mid-IR.
              }
         \label{sed_denis1620}
   \end{figure}

We measured the IRAC fluxes of DENIS162041 using a recommended aperture of 5 pixels, a sky annulus between 5 and 10 pixels, and aperture corrections of 1.061, 1.064, 1.067 and 1.08 for the four channels, respectively, as provided in the IRAC Data Handbook. We derive the final fluxes and their uncertainties by computing respectively the average and the deviation of the two consecutive measurements. The 5.8~$\mu$m flux has larger uncertainties because the object fell on the edge of a ditter pattern and the PSF of the object appears slightly degraded. In order to estimate an upper limit at 24~$\mu$m, we measured the flux within an aperture of 14 pixels ($\sim$35\arcsec) at nine positions centered around the expected location of the source, and derive the final upper limit from the 3-$\sigma$ standard deviation of the nine measurements. Table \ref{photom_denis1620} gives the photometry of this object, and Figure \ref{sed_denis1620} shows its SED.

\begin{deluxetable}{lllllllll}
\tablecaption{Photometry of the Targets\label{photom}}
\tablewidth{0pt}
\tablehead{
\colhead{Target} & \colhead{I} & \colhead{J} & \colhead{H} & \colhead{K} & \colhead{PAH1} & \colhead{SIV} & \colhead{PAH2} & \colhead{NeII}
}
\startdata
DENIS-PJ155556.0-204518 & 0.82 &  7.07  &  7.72 & 10.41 & 4.6$\pm$0.7 	& $<$6.5		& 2.4$\pm$1.0 	& \nodata \\
DENIS-PJ160334.7-182930 & 2.60 & 17.14  & 18.96 & 16.56 & 5.4$\pm$0.9  	& 6.3$\pm$3.4 	& 4.9$\pm$1.2 	& \nodata \\
DENIS-PJ160514.0-240652 & 1.83 & 12.89  & 12.15 & 10.36 & \nodata      	& \nodata 	& $<$3.2	 	& \nodata \\
DENIS-PJ160603.9-205644 & 0.63 &  6.38  &  7.04 &  7.00 & \nodata      	& \nodata 	& $<$6.3 	& \nodata \\
DENIS-PJ160951.1-272242 & 1.20 &  7.76  &  8.19 &  7.28 & \nodata 	& \nodata 	& $<$4.5 	& \nodata \\
DENIS-PJ160958.5-234518AB & 2.04 & 14.42  & 16.29 & 15.73 & \nodata 	& \nodata 	& 3.1$\pm$1.4 	& \nodata \\
DENIS-PJ161050.0-221252 & 2.15 & 14.15  & 15.00 & 13.69 & \nodata 	& \nodata 	& $<$2.7 	& \nodata \\
DENIS-PJ161816.2-261908AB & 3.26 & 21.95  & 30.47 & 27.11 & 2.9$\pm$0.9 	& \nodata 	& $<$29.1 	& \nodata \\
DENIS-PJ161833.2-251750AB & 2.19 & 16.52  & 21.22 & 20.13 & \nodata 	& \nodata 	& 2.4$\pm$1.0 	& \nodata \\
DENIS-PJ161939.8-214535AB & 1.24 &  9.47  &  9.90 &  9.17 & 6.9$\pm$1.0 	& 8.0$\pm$1.4 	& 9.6$\pm$1.2 	& $<$10.7   \\
\enddata
\tablecomments{I fluxes from DENIS magnitudes; J,H, and K fluxes from 2MASS magnitudes. All fluxes in mJy, 1-$\sigma$ uncertainties and 3-$\sigma$ upper limits derived as described in the text. DENIS161833AB was the only binary in the sample that could have been resolved with VISIR, but we detect only one of the components.}
\end{deluxetable}

\begin{deluxetable}{ll}
\tablecaption{Photometry of DENIS-PJ162041.5-242549.0 \label{photom_denis1620}}
\tablewidth{0pt}
\tablehead{
\colhead{Band} & \colhead{Flux} 
}
\startdata
I              & 0.26$\pm$0.02 \\
J              & 2.77$\pm$0.10 \\
H              & 4.39$\pm$0.12 \\
K              & 4.61$\pm$0.20 \\
3.6~$\mu$m     & 3.14$\pm$0.03 \\
4.5~$\mu$m     & 2.24$\pm$0.02 \\
5.8~$\mu$m     & 1.80$\pm$0.14 \\
8.0~$\mu$m     & 1.09$\pm$0.02 \\
24.0~$\mu$m    & $<$5.9 \\
\enddata
\tablecomments{I flux from DENIS magnitude; J,H, and K fluxes from 2MASS magnitudes. All fluxes in mJy, uncertainties and 3-$\sigma$ upper limits derived as described in the text.}
\end{deluxetable}

\section{Analysis \label{analysis}}

Figure \ref{seds} shows the spectral energy distributions (SED) of the VISIR targets, and compares to the relevant DUSTY atmospheric model SEDs from \citet{2001ApJ...556..357A} and $\lambda$-Ori M dwarfs from \citet{barrado_inprep} corrected for the distance of the association. For the comparison, we use the effective temperature scale derived for USco members by \citet{2004ApJ...609..885M}. This figure shows that three among the six targets detected in our mid-IR images display a significant excess (DENIS155556, DENIS160334,DENIS161939). Four sources (DENIS161833, DENIS161816, DENIS161050 and DENIS160958) show detections or upper limits very close to the photospheric values. Finally,  the three sources left (DENIS161603, DENIS160514,DENIS160951) were not detected and the derived  upper limits are so much larger than the photospheric values that we cannot draw any conclusion regarding the presense or absence of circumstellar disks around them.

Figure \ref{sed_denis1620} shows the SED of DENIS162041. There is a clear discrepency between the SED and the DUSTY model/comparison M dwarf SEDs. The fluxes are lower than expected, with a difference increasing toward bluer wavelengths, suggesting that the discrepency might be due to reddening. The Spitzer images shows that DENIS162041 is indeed located in an area where extinction is relatively high, as shown in Figure \ref{denis1620_spitzer}, and a much better fit of the SED is obtained when unreddening the SED with $A_{V}$=3.3, assuming a standard reddening law as given in \citet{1985ApJ...288..618R}. In any case DENIS162041 does not display any mid-IR excess.

\subsection{Inner disks and multiple systems}
Recent studies have found a significant fraction of very low mass stars and brown dwarfs multiple systems in star forming regions \citep[between 9--33\% for separations greater than 4~AU in Taurus and USco, ][]{2006A&A...451..177B,2006astro.ph..2449K,2005ApJ...633..452K}, and several young multiple systems exhibit evidences for disks and/or accretion \citep[][]{2005prpl.conf.8329C,2004A&A...424..213B,2004A&A...427..245S}. As in the case of stars, the formation of multiple systems surrounded by disk(s) seems to be a relatively common outcome of sub-stellar formation. Among the four known multiple systems detected in our VISIR images, only DENIS161939AB shows a clear mid-IR excess indicating the presence of an inner disk. This binary has a separation of 0\farcs177 \citep{2006A&A...451..177B}, or 26~AU at the distance of USco, and the unresolved system was classified as an accretor by \citet{2004AJ....127..449M}. DENIS161833AB was the only binary in the sample that could have been resolved with VISIR but we only detect one of the members.

Two targets (DENIS161050 and DENIS160334) are brighter by a factor of 2 than the atmospheric models with effective temperature corresponding to their spectral types. Two explanations can be given: either the targets are unresolved binaries, and/or they lay closer to the Sun than the rest of the targets. Figure \ref{seds} shows that the fit becomes very good if one simply considers that these objects are unresolved binaries, and compare their SED with the sum of two atmospheric models of the same temperature. This assumes that the two components would have similar effective temperatures, as suggested by the results of \citet{2006A&A...451..177B} showing that all resolved very low mass and BD USco binaries have close to equal mass ratios.  These objects were not resolved in the high angular resolution observations of \citet{2006A&A...451..177B}. If multiples, the companions must therefore be at separations $<$5~AU and/or $\Delta$Mag(Ks)$>$2~mag. The data could be equally well fitted with a single theoretical SED only $\sqrt{2}$ closer to the Sun than the rest of the sample (which would place them at $\sim$100~pc). \citet{1998AJ....115..351M} indeed showed that members of the USco association can be found as close as 80~pc from the Sun. 

In either case, the SED of DENIS161050 do not show any significant excess, while that of DENIS160334 displays a substantial excess at wavelengths greater than 8~$\mu$m.

\subsection{Disk Fraction}
Three objects (DENIS155556, DENIS160334, and DENIS161939AB) among the six with mid-IR detections (the latter three plus DENIS160958, DENIS161816AB, and DENIS161833AB) show a clear excess at 8~$\mu$m  and above. DENIS162041 was detected in the Spitzer images but would not have been detected in our ground-based survey. We therefore do not include it in the following discussion. The upper limit on DENIS161050 mid-IR flux is only a factor of $<$2 above the atmosphere, and the presence of disk can be ruled out at a relatively high confidence level. \citet{2004AJ....127..449M} classified DENIS160603 as an accretor on the basis of its strong H$\alpha$ emission ($W(H\alpha)$=105$\pm$12~\AA). We do not detect this object in our mid-IR images but the upper limit on its mid-IR flux is too high to put any valuable constraints on the presence or absence of circumstellar material. If we assume that we did not detect this object because of the limited sensitivity of our images, we can add it to the statistics, and obtain a disk frequency\footnote{Deriving small number statistics, we estimate the uncertainties by constructing a probability distribution for the disk frequency given the total sample size and the number of disks in the sample.} of 50\%$\pm$16\% (4 disks over 8 targets). The remaining two upper limits (for DENIS160514 and DENIS160951) are too high to be conclusive and we do not include these objects in the statistics. 

The disk frequency estimated above is consistent with that reported by \citet{2003AJ....126.1515J} using L-band photometry  (4 disks among 8 targets, 50\%$\pm$16\%, with uncertainties recomputed as described previously). Since their sample does not overlap with the one presented here, and although their study was performed using a different technique, one can tentatively add the two samples to derive an improved disk fraction of 50\%$\pm$12\% (8/16). This result is comparable to the one reported for Scorpius-Centaurus solar-mass stars, where \citet{2005ApJ...623..493C} recently reported a preliminary estimate of the disk fraction among F and G stars ($>$35\%) using Spitzer photometry.

\subsection{Sub-stellar formation, disk evolution and multiple systems}

Table \ref{bindisk} summarizes the properties of multiplicity, disk and accretion of the sample. The 10$\mu$m excesses we observe are produced by rather warm dust (300~K), close to the object ($<$1~AU). We do not probe any outer disk, and cannot say anything about the presence of circumbinary material that would be located at distances larger than the separations of these systems. We cannot either tell whether only one or the two components have a disk, but this should not change our estimate of the disk fraction since binaries are counted as single objects anyway, assuming that the two components must be coeval. 

Note also that these results consider that the inner disk frequency of multiple systems and single objects at the age of Upper Scorpius must be the same, or, in other words, that the presence of a companion at such separations does not affect the evolution of the disk(s) to the point that we would not have detected it. This assumption seems reasonable considering the relatively large separations of these binaries \citep[12, 21, 26 and 134~AU, see Table \ref{bindisk} and ][]{2006A&A...451..177B} compared to the typical sizes of BD disks to which we are sensitive with these mid-IR observations \citep[1~AU, see e.g ][ for a discussion about disk evolution in young binaries]{2006astro.ph..4031M,2003A&A...409L..25M}.

\subsection{Disks and Accretion}
Only one (DENIS161939) of the two objects of our sample classified as accretors by \citet{2004AJ....127..449M} is detected in our mid-IR images. As explained above, the mid-IR upper limit derived for the other one (DENIS160603) is inconclusive. The accretor frequency (16.3\%$\pm$6.2\%) reported by \citet{2004AJ....127..449M} seems to be substantially lower than the disk frequency at the age of USco, indicating that transition disks are much more numerous than accretion disks, which is consistent with the timescale of disk decay observed for young ultracool dwarf disks \citep{2004A&A...427..245S,2005MmSAI..76..303M}. 

While the dust contributes to a small fraction of the mass of the disk ($\sim$1\%), it is the most essential ingredient governing the heating/cooling of the disk, and therefore ultimately its physical and geometrical evolution. The dust consists mainly of silicate grains with strong emission features in the mid-IR around 9.3~$\mu$m and 11.3~$\mu$m. In their recent study, \citet{2005Sci...310..834A} show that these two silicate features provide unique diagnostics of the grain growth, crystallization and dust settling in the disk. Unfortunately, because of the limited sampling of the SEDs and because of the coupled influence of grain sizes and dust geometry on the SED, the results we present here do not allow us to perform a detailed analysis of the disk geometry and grain sizes. Complementary observations with e.g Spitzer are required before a more detailed comparison of the SEDs of the targets with mid-IR excesses to disk models can be done, but the current results nevertheless show that some of the Upper Scorpius ultracool dwarfs are undergoing similar stages of evolution than the Cha~I objects observed by \citet{2005Sci...310..834A}.

\section{Concluding Remarks}
We report mid-IR observations of 11 ultracool dwarfs members of the Upper Scorpius association. Seven of them are detected in at least one of the mid-IR images, and 3 show a clear excess indicating the presence of circumstellar material. The relatively high frequency of disks (50$\pm$12\%) found among Upper Scorpius ultracool dwarfs (5~Myr) is comparable to those reported by e.g \citet{2003AJ....126.1515J, 2005ApJ...631L..69L} for a variety of younger associations (1--2~Myr), possibly indicating that the timescales of disk decay are longer for ultracool dwarfs than for solar-mass stars. Although the sample we present is too small to draw any statistically meaningful conclusion, there seems to be no significant difference between the disk frequency of single and multiple ultracool dwarfs. Since these relatively wide systems have most likely formed like stars (they are indeed very unlikely to have survived a gravitational ejection), this similarity between single and multiple objects might be an additional clue that most ultracool dwarfs must form like stars rather than ejected embryos. Further studies on larger samples of single and multiple systems are required to confirm these observations. Finally, only one of the objects displaying mid-IR excess also shows evidence of accretion, indicating that the frequency of transition disks in Upper Scorpius must be signiticantly higher than that of accretion disks.

\begin{acknowledgements}
We thank our referee, Kevin Luhman, for his prompt review and interesting comments which helped improving the manuscript.

H. Bouy thanks the telescope operators and the support astronomers, as well as the whole Paranal SciOps team for their very nice and efficient support. H. Bouy acknowledges the funding from the European Commission's Sixth Framework Program as a Marie Curie Outgoing International Fellow (MOIF-CT-2005-8389).

This work is based on observations collected at the European Southern Observatory (Paranal, Chile), program 075.C-0148, P.I. Bouy. This work has made use of the CDS VizieR on-line tools \citep{Vizier}. This work makes use of DENIS data. The DENIS project has been partly funded by the SCIENCE and the HCM plans of the European Commission under grants CT920791 and CT940627. It is supported by INSU, MEN and CNRS in France, by the State of Baden-W\"urttemberg in Germany, by DGICYT in Spain, by CNR in Italy, by FFwFBWF in Austria, by FAPESP in Brazil, by OTKA grants F-4239 and F-013990 in Hungary, and by the ESO C\&EE grant A-04-046. This publication makes use of data products from the Two Micron All Sky Survey, which is a joint project of the University of Massachusetts and the Infrared Processing and Analysis Center/California Institute of Technology, funded by the National Aeronautics and Space Administration and the National Science Foundation.

\end{acknowledgements}

\end{document}